\documentclass[11pt]{article}
\usepackage{amsfonts}
\usepackage{graphicx}
\usepackage{amsmath}
\usepackage{amssymb}
\usepackage{caption2}
\begin{document}
\bibliographystyle{prsty}
\begin{center}
{\large {\bf \sc{  Analysis of the scalar doubly
heavy tetraquark states with QCD sum rules }}} \\[2mm]
Zhi-Gang Wang\footnote{E-mail,wangzgyiti@yahoo.com.cn.  }, Yan-Mei Xu, Hui-Juan Wang      \\
 Department of Physics, North China Electric Power University,
Baoding 071003, P. R. China
\end{center}

\begin{abstract}
In this article, we  perform a systematic study of  the mass
spectrum of the scalar doubly charmed  and doubly bottom tetraquark
states using the QCD sum rules.
\end{abstract}

 PACS number: 12.39.Mk, 12.38.Lg

Key words: Tetraquark state, QCD sum rules

\section{Introduction}

  The $Z(4430)$ and  the $Z(4050)$,
$Z(4250)$ observed in the  decay modes $\psi^\prime\pi^+$ and
$\chi_{c1}\pi^+$ respectively  by the
 Belle collaboration are the most
interesting subjects \cite{Belle-z4430,Belle-z4430-PRD,Belle-chipi}.
We can distinguish the multiquark states
 from the hybrids or charmonia with the criterion of
non-zero charge. They can't be pure $c\bar{c}$ states due to the
positive charge,  and must be some special combinations of the
 $c\bar{c}u\bar{d}$ tetraquark states, irrespective of the molecule  type and the
diquark-antidiquark type. If those states are confirmed in the
future, they are excellent candidates for the heavy tetraquark
states of the $Qq\bar{Q}\bar{q}'$ type. It is interesting to explore
the possibility that  whether or not there exist doubly heavy
tetraquark states of the $QQ\bar{q}\bar{q}'$ type.

On the other hand, the QCD sum rules is a powerful theoretical tool
in studying the ground state hadrons \cite{SVZ79,Reinders85}. In the
QCD sum rules, the operator product expansion is used to expand the
time-ordered currents into a series of quark and gluon condensates
which parameterize the long distance properties of the QCD vacuum.
Based on the quark-hadron duality, we can obtain copious information
about the hadronic parameters at the phenomenological side
\cite{SVZ79,Reinders85}.

There have been several successful applications of the QCD sum rules
in studying the hidden charmed and hidden bottom tetraquark states
($Qq\bar{Q}\bar{q}'$ type). In
Refs.\cite{Wang08072,WangScalar,WangScalar-2}, we study the mass
spectrum of the scalar hidden charmed and hidden bottom tetraquark
states in a systematic way using the QCD sum rules, and identify the
$Z(4250)$ tentatively as a scalar tetraquark state of the
diquark-antidiquark type; while in Ref.\cite{Lee-4250} the
 $Z(4050)$ and $Z(4250)$ are interpreted as  the $D_1\bar{D}$ molecular state. In
Refs.\cite{Wang0807,WangVector}, we study the mass spectrum of the
vector hidden charmed and hidden bottom tetraquark states
systematically using the QCD sum rules. In Ref.\cite{WangAxial}, we
perform a systematic study of the mass spectrum of the axial-vector
hidden charmed and hidden bottom tetraquark states using the QCD sum
rules, and identify the $Z(4430)$ tentatively as an axial-vector
tetraquark state of the diquark-antidiquark type. In
Refs.\cite{Lee-1, Lee-2}, Lee et al study the $Z(4430)$ with the QCD
sum rules and observe that the $Z(4430)$ maybe a  $0^-$ molecular
type or diquark-antidiquark type tetraquark state.
 In Ref.\cite{Zhu2010}, Chen et al study the $0^{--}$ hidden charmed and hidden bottom
tetraquark states in details  with the QCD sum rules.

In Ref.\cite{Navarra-QQ}, Navarra et al use the QCD sum rules to
study the possible existence of the doubly heavy tetraquark states
$QQ\bar{u}\bar{d}$ with $J^{P}=1^{+}$. There have been several other
theoretical approaches in studying  the doubly heavy tetraquark
states, such as the potential  models and QCD inspired potential
models \cite{Richard-1,Richard-2,Richard-3}, solving the four-body
problem within a non-relativistic quark model
\cite{Silvestre-Brac-1993,Janc-2004}, the variational method
combined with a non-relativistic potential model \cite{Brink-1998},
the chiral constituent quark model \cite{Vijande-2004,Vijande-2006},
the  semi-empirical mass relations \cite{Gelman-2003}, the
relativistic quark model based on a quasipotential approach in QCD
\cite{Ebert-2007}, etc.
 Whether or not there exist the doubly charmed  or doubly bottom
tetraquark configurations is of great importance itself, because it
provides a new opportunity for a deeper understanding of the low
energy QCD.

It is interesting to study the mass spectrum of the
 doubly heavy  tetraquark states ($QQ\bar{q}\bar{q}'$ type) with the QCD sum rules, and
make an independent estimation from QCD. In
Refs.\cite{Wang-QQ1,Wang-QQ2,Wang-QQ3}, we study the mass spectrum
of the ${1\over 2}^\pm$ and ${3\over2}^\pm$ doubly heavy baryon
states ($QQq$ type) in a systematic way using the QCD sum rules. In
this article, we extend our previous works to study the mass
spectrum of the scalar doubly charmed and doubly bottom tetraquark
states in a systematic way with the QCD sum rules.

We take the diquarks as the basic constituents   following  Jaffe
and Wilczek \cite{Jaffe2003,Jaffe2004}, and construct the doubly
heavy tetraquark states with the diquark and antidiquark pairs. The
diquarks have five Dirac tensor structures, scalar $C\gamma_5$,
pseudoscalar $C$, vector $C\gamma_\mu \gamma_5$, axial-vector
$C\gamma_\mu $  and tensor $C\sigma_{\mu\nu}$, where  $C$ is the
charge conjunction matrix. The structures $C\gamma_\mu $ and
$C\sigma_{\mu\nu}$ are symmetric while  the structures $C\gamma_5$,
$C$ and $C\gamma_\mu \gamma_5$ are antisymmetric.

The scattering amplitude for one-gluon exchange in an $SU(N_c)$
gauge theory is proportional to
\begin{eqnarray}
T^a_{ki}
T^a_{lj}&=&-\frac{N_c+1}{4N_c}(\delta_{jk}\delta_{il}-\delta_{ik}\delta_{jl})
 +\frac{N_c-1}{4N_c}(\delta_{jk}\delta_{il}+\delta_{ik}\delta_{jl})\, ,
\end{eqnarray}
where the $T^a$ is the generator of the gauge group, and the $i,j$
and $k,l$ are the  color indexes  of the two quarks in the incoming
and outgoing channels respectively.   For $N_c=3$, the negative sign
in front of the antisymmetric  antitriplet indicates the interaction
is attractive, while the positive sign in front of the symmetric
sextet indicates
 the interaction   is repulsive \cite{Huang-2005}. On the other
 hand, the scattering amplitude for one-gluon exchange in the Dirac
 spinor space is proportional to
 \begin{eqnarray}
 (\gamma_\mu)_{ij}
 (\gamma^\mu)_{kl}&=&-(\gamma_5C)_{ik}(C\gamma_5)_{lj}+(C)_{ik}(C)_{lj}+\frac{1}{2}(\gamma_5\gamma_\alpha C)_{ik}(C\gamma^\alpha\gamma_5)_{lj}\nonumber\\
 &&-\frac{1}{2}( \gamma_\alpha C)_{ik}(C\gamma^\alpha )_{lj} \, ,
 \end{eqnarray}
 the negative sign in front of the scalar and axial-vector channels
 indicates  the interaction is attractive.

 For the doubly heavy quark system with the same flavor $Q^i C\Gamma Q^j $,
 where the $\Gamma$ denote the Dirac matrixes $1$, $\gamma_5$,  $\gamma_\mu $, $\gamma_\mu \gamma_5$
   and  $\sigma_{\mu\nu}$, the color indexes $i$ and $j$ should be
   antisymmetric, i.e.
   \begin{eqnarray}
Q^i C\Gamma Q^j \sim  \epsilon_{ijk} Q^i C\Gamma Q^j \, .
\end{eqnarray}
In the case of the antisymmetric structures $C\gamma_5$, $C$ and
$C\gamma_\mu \gamma_5$, the fermi statistics forbids  the
formulation of the diquark states.

In this article, we use the symmetric structure $C\gamma_\mu $ to
construct the interpolating currents $J(x)$ to study the doubly
charmed  and doubly bottom tetraquark states $Z$:
\begin{eqnarray}
J_{qq}(x)&=& \epsilon^{ijk}\epsilon^{imn}Q_j^T(x) C\gamma_\mu Q_k(x)\bar{q}_m(x)  \gamma^\mu C \bar{q}_n^T(x)\, , \nonumber\\
J_{qs}(x)&=& \epsilon^{ijk}\epsilon^{imn}Q_j^T(x) C\gamma_\mu Q_k(x)\bar{q}_m(x)  \gamma^\mu C \bar{s}_n^T(x)\, , \nonumber\\
 J_{ss}(x)&=& \epsilon^{ijk}\epsilon^{imn}Q_j^T(x) C\gamma_\mu Q_k(x)\bar{s}_m(x)  \gamma^\mu C \bar{s}_n^T(x)\, ,
\end{eqnarray}
where $q=u,d$. In the isospin limit, the interpolating currents
$J(x)$ result in three distinct expressions for the correlation
functions $\Pi(p)$, which are characterized by the number of the $s$
quark they contain. In Refs.\cite{WangScalar,WangVector}, we observe
that the ground state masses of the scalar and vector hidden charmed
and hidden bottom tetraquarks are characterized by the number of the
$s$ quarks they contain, $M_{0}\leq M_{s}\leq M_{ss}$; the energy
gap between $M_{0}$ and $M_{ss}$ is about $(0.05-0.15)\,\rm{GeV}$.
In this article, we study the interpolating currents which contains
zero and two $\bar{s}$ quarks for simplicity.

Lattice QCD calculations for the light flavors  indicate that the
strong attraction in the scalar diquark channels favors the
formation of  good diquarks, the weaker attraction (the quark-quark
correlation is rather weak) in the axial-vector diquark channels
maybe form bad diquarks, the energy gap between the axial-vector and
scalar diquarks is about $\frac{2}{3}$ of the $\Delta$-nucleon mass
splitting, i.e. $\approx 0.2\,\rm{GeV}$ \cite{Latt-1,Latt-2}, which
is expected from the hypersplitting color-spin  interaction
$\frac{1}{m_im_j}\vec{T}_{i}\cdot \vec{T}_{j} \vec{\sigma}_i \cdot
\vec{\sigma}_j$ \cite{Jaffe2004}. The coupled rainbow
Dyson-Schwinger equation and ladder Bethe-Salpeter equation also
indicate such an energy hierarchy \cite{BS-diquark}. Comparing with
the spin independent term $\vec{T}_{i}\cdot \vec{T}_{j}$, the
contribution from the hypersplitting color-spin  interaction
$\frac{1}{m_im_j}\vec{T}_{i}\cdot \vec{T}_{j} \vec{\sigma}_i \cdot
\vec{\sigma}_j$ is greatly suppressed by the inverse constituent
quark masses. It is  possible to form axial-vector diquark states,
although the hypersplitting color-spin interaction
$\frac{1}{m_im_j}\vec{T}_{i}\cdot \vec{T}_{j} \vec{\sigma}_i \cdot
\vec{\sigma}_j$ is repulsive in this channel. If we take the scalar
light diquark states as the basic constituents, additional relative
$P$-waves are needed to obtain  the correct zero spin. In the
conventional quark models, additional $P$-wave excitation costs
about $0.5\,\rm{GeV}$, the ground states should be constructed with
the axial-vector antidiquark states.

The article is arranged as follows:  we derive the QCD sum rules for
  the scalar doubly charmed  and doubly bottom tetraquark states  $Z$  in Sect.2; in Sect.3, we present the
 numerical results and discussions; and Sect.4 is reserved for our
conclusions.

\section{QCD sum rules for  the scalar  tetraquark states $Z$ }
In the following, we write down  the two-point correlation functions
$\Pi(p)$  in the QCD sum rules,
\begin{eqnarray}
\Pi(p)&=&i\int d^4x e^{ip \cdot x} \langle
0|T\left[J(x)J^{\dagger}(0)\right]|0\rangle \, ,
\end{eqnarray}
where the  $J(x)$  denotes the interpolating currents $J_{qq}(x)$
and $J_{ss}(x)$.

We can insert  a complete set of intermediate hadronic states with
the same quantum numbers as the current operators $J(x)$  into the
correlation functions  $\Pi(p)$  to obtain the hadronic
representation \cite{SVZ79,Reinders85}. After isolating the ground
state contribution from the pole term of the $Z$, we get the
following result,
\begin{eqnarray}
\Pi(p)&=&\frac{\lambda_{Z}^2}{M_{Z}^2-p^2} +\cdots \, \, ,
\end{eqnarray}
where the pole residue (or coupling) $\lambda_Z$ is defined by
\begin{eqnarray}
\lambda_{Z}   &=& \langle 0|J(0)|Z(p)\rangle \, .
\end{eqnarray}

 After performing the standard procedure of the QCD sum rules, we obtain the following  two sum rules:
\begin{eqnarray}
\lambda_{Z}^2 e^{-\frac{M_{Z}^2}{M^2}}= \int_{\Delta_{Z}}^{s^0_{Z}}
ds \rho_Z(s)e^{-\frac{s}{M^2}} \, ,
\end{eqnarray}
  the explicit expressions of the spectral densities $\rho_Z(s)$
are  presented in  the appendix,  the $s^0_Z$ is the continuum
threshold parameter and the $M^2$ is the Borel  parameter.
       We can obtain  two   sum rules in  the
$cc\bar{q}\bar{q}$ and $bb\bar{q}\bar{q}$ channels with a simple
replacement $m_s\rightarrow m_q$,  $\langle
\bar{s}s\rangle\rightarrow\langle \bar{q}q\rangle$ and $\langle
\bar{s}g_s \sigma Gs\rangle\rightarrow\langle \bar{q}g_s \sigma
Gq\rangle$.

 We carry out the operator
product expansion to the vacuum condensates adding up to
dimension-10. In calculation, we
 take   vacuum saturation for the high
dimension vacuum condensates, they  are always
 factorized to lower condensates with vacuum saturation in the QCD sum rules,
  factorization works well in  large $N_c$ limit. In reality, $N_c=3$, some  ambiguities may come from
the vacuum saturation assumption.

We take into account the contributions from the quark condensates,
mixed condensates, and neglect the contributions from the gluon
condensate. The gluon condensate $\langle
\frac{\alpha_sGG}{\pi}\rangle$  is of higher order in $\alpha_s$,
and its contributions   are suppressed by  very large denominators
and would not play any significant role for the light tetraquark
states \cite{Wang1,Wang2}, the heavy tetraquark state
\cite{Wang08072} and the  heavy molecular states
\cite{Wang0904,Wang0907}.

In the special case of the $Y(4660)$ (as a $\psi'f_0(980)$ bound
state) and its pseudoscalar partner $\eta_c'f_0(980)$, the
contributions from the gluon condensate $\langle \frac{\alpha_s
GG}{\pi} \rangle $ are rather large \cite{WangZhang1,WangZhang2}. If
we take a simple replacement $\bar{s}(x)s(x)\rightarrow \langle
\bar{s}s\rangle$ and $\left[\bar{u}(x)u(x)+\bar{d}(x)d(x)
\right]\rightarrow 2\langle\bar{q}q\rangle$ in the interpolating
currents,  the standard  heavy quark currents $Q(x)\gamma_\mu Q(x)$
and $Q(x)i\gamma_5 Q(x)$ are obtained, where the gluon condensate
$\langle \frac{\alpha_s GG}{\pi} \rangle $ plays an important rule
in the QCD sum rules \cite{SVZ79}. The interpolating currents
constructed from the diquark-antidiquark pairs do not have such
feature.

We also neglect the terms proportional to the $m_u$ and $m_d$, their
contributions are of minor importance due to the small values of the
$u$ and $d$ quark masses.

 Differentiating  the Eq.(8) with respect to  $\frac{1}{M^2}$, then eliminate the
 pole residues $\lambda_{Z}$, we can obtain the sum rules for
 the masses  of the $Z$,
 \begin{eqnarray}
 M_{Z}^2= \frac{\int_{\Delta_{Z}}^{s^0_{Z}} ds \frac{d}{d(-1/M^2)}
\rho_Z(s)e^{-\frac{s}{M^2}} }{\int_{\Delta_{Z}}^{s^0_{Z}} ds
\rho_Z(s)e^{-\frac{s}{M^2}}}\, .
\end{eqnarray}

\section{Numerical results and discussions}
The input parameters are taken to be the standard values $\langle
\bar{q}q \rangle=-(0.24\pm 0.01 \,\rm{GeV})^3$, $\langle \bar{s}s
\rangle=(0.8\pm 0.2 )\langle \bar{q}q \rangle$, $\langle
\bar{q}g_s\sigma Gq \rangle=m_0^2\langle \bar{q}q \rangle$, $\langle
\bar{s}g_s\sigma Gs \rangle=m_0^2\langle \bar{s}s \rangle$,
$m_0^2=(0.8 \pm 0.2)\,\rm{GeV}^2$,  $m_s=(0.14\pm0.01)\,\rm{GeV}$,
$m_c=(1.35\pm0.10)\,\rm{GeV}$ and $m_b=(4.8\pm0.1)\,\rm{GeV}$ at the
energy scale  $\mu=1\, \rm{GeV}$ \cite{SVZ79,Reinders85,Ioffe2005}.

The $Q$-quark masses appearing in the perturbative terms  are
usually taken to be the pole masses in the QCD sum rules, while the
choice of the $m_Q$ in the leading-order coefficients of the
higher-dimensional terms is arbitrary \cite{NarisonBook,Kho9801}.
The $\overline{MS}$ mass $m_c(m_c^2)$ relates with the pole mass
$\hat{m}_c$ through the relation $ m_c(m_c^2)
=\hat{m}_c\left[1+\frac{C_F \alpha_s(m_c^2)}{\pi}+\cdots\right]^{-1}
$. In this article, we take the approximation
$m_c(m_c^2)\approx\hat{m}_c$ without the $\alpha_s$ corrections for
consistency. The value listed in the Particle Data Group is
$m_c(m_c^2)=1.27^{+0.07}_{-0.11} \, \rm{GeV}$ \cite{PDG}, it is
reasonable to take
$\hat{m}_c=m_c(1\,\rm{GeV}^2)=(1.35\pm0.10)\,\rm{GeV}$. For the $b$
quark,  the $\overline{MS}$ mass
$m_b(m_b^2)=4.20^{+0.17}_{-0.07}\,\rm{GeV}$ \cite{PDG}, the
  gap between the energy scale $\mu=4.2\,\rm{GeV}$ and
 $1\,\rm{GeV}$ is rather large, the approximation $\hat{m}_b\approx m_b(m_b^2)\approx m_b(1\,\rm{GeV}^2)$ seems rather crude.
  It would be better to understand the quark masses $m_c$ and $m_b$ we
take at the energy scale $\mu^2=1\,\rm{GeV}^2$ as the effective
quark masses (or just the mass parameters). Our previous works on
the mass spectrum of the heavy and doubly heavy baryon states
indicate such parameters can lead to satisfactory results
\cite{Wang-QQ1,Wang-QQ2,Wang-QQ3,Wang-H-1,Wang-H-2}.

In calculation, we  also neglect  the contributions from the
perturbative corrections.  Those perturbative corrections can be
taken into account in the leading logarithmic
 approximations through  anomalous dimension factors. After the Borel transform, the effects of those
 corrections are  to multiply each term on the operator product
 expansion side by the factor, $ \left[ \frac{\alpha_s(M^2)}{\alpha_s(\mu^2)}\right]^{2\Gamma_{J}-\Gamma_{\mathcal
 {O}_n}}  $,
 where the $\Gamma_{J}$ is the anomalous dimension of the
 interpolating current $J(x)$ and the $\Gamma_{\mathcal {O}_n}$ is the anomalous dimension of
 the local operator $\mathcal {O}_n(0)$. We carry out the operator product expansion at a special energy
scale $\mu^2=1\,\rm{GeV}^2$, and  set the factor $\left[
\frac{\alpha_s(M^2)}{\alpha_s(\mu^2)}\right]^{2\Gamma_{J}-\Gamma_{\mathcal
{O}_n}}\approx1$, such an approximation maybe result in some scale
dependence  and  weaken the prediction ability. In this article, we
study the scalar doubly charmed  and doubly bottom tetraquark states
systemically, the predictions are still robust as we take the
analogous criteria in those sum rules.

In the conventional QCD sum rules \cite{SVZ79,Reinders85}, there are
two criteria (pole dominance and convergence of the operator product
expansion) for choosing  the Borel parameter $M^2$ and threshold
parameter $s_0$. We impose the two criteria on the scalar doubly
heavy tetraquark states to choose the Borel parameter $M^2$ and
threshold parameter $s_0$.

The vacuum condensates of the high  dimension
 play an important role in choosing the Borel parameter $M^2$.
The condensate of the highest dimension $\langle \bar{s}g_s \sigma G
s\rangle^2$ is counted as $\mathcal {O}(\frac{m_Q^2}{M^2})$,
$\mathcal {O}(\frac{m_Q^4}{M^4})$ or $\mathcal
{O}(\frac{m_Q^6}{M^6})$, and the corresponding contributions are
greatly enhanced  at small $M^2$, and result in rather  bad
convergent behavior in the operator product expansion, we have to
choose large Borel parameter $M^2$. We insist on taking into account
the high dimensional  vacuum condensates, as the interpolating
current consists  of  a (heavy)diquark-(light)antidiquark pair, one
of the highest dimensional vacuum condensates is $\langle
\bar{s}s\rangle^2 \langle \frac{\alpha_s GG}{\pi}\rangle$, we have
to take into account the condensate $\langle \bar{s}g_s \sigma G
s\rangle^2$ for consistence.

 The contributions from the high
dimension vacuum condensates  in the operator product expansion are
shown in Fig.1, where (and thereafter) we  use the
$\langle\bar{s}s\rangle$ to denote the quark condensates
$\langle\bar{q}q\rangle$, $\langle\bar{s}s\rangle$ and the
$\langle\bar{s}g_s \sigma Gs\rangle$ to denote the mixed condensates
$\langle\bar{q}g_s \sigma Gq\rangle$, $\langle\bar{s}g_s \sigma
Gs\rangle$. From the figures, we can see that  the contributions
from the high dimension condensates are very large and change
quickly with variation of the Borel parameter at the values $M^2\leq
2.6 \,\rm{GeV}^2$ and $M^2\leq 7.2 \,\rm{GeV}^2$ in the doubly
charmed  and doubly bottom channels respectively, such an unstable
behavior cannot lead to stable sum rules, our numerical results
confirm this conjecture, see Fig.3.

At the values $M^2\geq 2.6\,\rm{GeV}^2 $ and $s_0\geq
25\,\rm{GeV}^2,\,24\,\rm{GeV}^2$, the contributions from the
$\langle \bar{s}s\rangle^2+\langle \bar{s}s\rangle \langle
\bar{s}g_s \sigma Gs\rangle $ term are less than $9\%,\,23\%$ in the
channels $cc\bar{s}\bar{s}$, $cc\bar{q}\bar{q}$
 respectively; the contributions from the vacuum
condensate of the highest dimension $\langle\bar{s}g_s \sigma
Gs\rangle^2$ are less than $4\%,\,5\%$ in the channels
$cc\bar{s}\bar{s}$, $cc\bar{q}\bar{q}$ respectively; we expect the
operator product expansion is convergent in the doubly charmed
channels.

At the values $M^2\geq 7.2\,\rm{GeV}^2 $ and $s_0\geq
140\,\rm{GeV}^2,\,138\,\rm{GeV}^2$, the contributions from the
$\langle \bar{s}s\rangle^2+\langle \bar{s}s\rangle \langle
\bar{s}g_s \sigma Gs\rangle $ term are less than $6\%,\,16\%$ in the
channels $bb\bar{s}\bar{s}$, $bb\bar{q}\bar{q}$ respectively; the
contributions from the vacuum condensate of the highest dimension
$\langle\bar{s}g_s \sigma Gs\rangle^2$ are less than $5\%,\,8\%$ in
the channels $bb\bar{s}\bar{s}$, $bb\bar{q}\bar{q}$ respectively; we
expect the operator product expansion is convergent in the doubly
bottom channels.

 In this article, we take the uniform Borel parameter
$M^2_{min}$, i.e. $M^2_{min}\geq 2.6 \, \rm{GeV}^2$  and
$M^2_{min}\geq 7.2 \, \rm{GeV}^2$ in the doubly charmed  and doubly
bottom channels respectively.

In Fig.2, we show the  contributions from the pole terms with
variation of the Borel parameters and the threshold parameters. The
pole contributions are larger than (or equal) $50\%,\,47\%$ at the
value $M^2 \leq 3.3 \, \rm{GeV}^2 $ and $s_0\geq
25\,\rm{GeV}^2,\,24\,\rm{GeV}^2$
 in the channels $cc\bar{s}\bar{s}$, $cc\bar{q}\bar{q}$
 respectively, and larger than (or equal)
$52\%,\,50\%$ at the value $M^2 \leq 8.2 \, \rm{GeV}^2 $ and
$s_0\geq 140\,\rm{GeV}^2,\,138\,\rm{GeV}^2$
 in  the channels $bb\bar{s}\bar{s}$, $bb\bar{q}\bar{q}$ respectively. Again we take the uniform Borel
parameter $M^2_{max}$, i.e. $M^2_{max}\leq 3.3 \, \rm{GeV}^2$ and
$M^2_{max}\leq 8.2 \, \rm{GeV}^2$ in the doubly charmed and doubly
bottom channels respectively.

In this article, the threshold parameters are taken as
$s_0=(26\pm1)\,\rm{GeV}^2$, $(25\pm1)\,\rm{GeV}^2$,
$(142\pm2)\,\rm{GeV}^2$, $(140\pm2)\,\rm{GeV}^2$ in the channels
$cc\bar{s}\bar{s}$, $cc\bar{q}\bar{q}$,
    $bb\bar{s}\bar{s}$, $bb\bar{q}\bar{q}$ respectively;
   the Borel parameters are taken as $M^2=(2.6-3.3)\,\rm{GeV}^2$ and
   $(7.2-8.2)\,\rm{GeV}^2$ in the
doubly charmed  and doubly bottom channels respectively.
      In those regions,  the pole contributions are about
$(50-80)\%$, $(47-78)\%$,  $(52-71)\%$, $(50-70)\%$  in the channels
$cc\bar{s}\bar{s}$, $cc\bar{q}\bar{q}$,
    $bb\bar{s}\bar{s}$, $bb\bar{q}\bar{q}$  respectively;   the two criteria of the QCD sum rules
are fully  satisfied  \cite{SVZ79,Reinders85}.

From Fig.2, we can see that the Borel windows $M_{max}^2-M_{min}^2$
change with variations of the  threshold parameters $s_0$. In this
article, the Borel windows  are taken as $0.7\,\rm{GeV}^2$ and
$1.0\,\rm{GeV}^2$ in the doubly charmed  and doubly bottom channels
respectively; they are small enough.  If we take larger threshold
parameters,  the Borel windows are larger and the resulting  masses
are larger, see Fig.3. In this article, we intend to  calculate the
possibly  lowest masses which are supposed to be the ground state
masses  by imposing the two criteria of the QCD sum rules.

Taking into account all uncertainties of the relevant  parameters,
finally we obtain the values of the masses and pole resides of
 the scalar doubly heavy tetraquark states  $Z$, which are  shown in Figs.4-5 and Table 1.

In this article,  we calculate the uncertainties $\delta$  with the
formula
\begin{eqnarray}
\delta=\sqrt{\sum_i\left(\frac{\partial f}{\partial
x_i}\right)^2\mid_{x_i=\bar{x}_i} (x_i-\bar{x}_i)^2}\,  ,
\end{eqnarray}
 where the $f$ denote  the
hadron mass  $M_Z$ and the pole residue $\lambda_Z$,  the $x_i$
denote the relevant parameters $m_c$, $m_b$, $\langle \bar{q}q
\rangle$, $\langle \bar{s}s \rangle$, $\cdots$. As the partial
 derivatives   $\frac{\partial f}{\partial x_i}$ are difficult to carry
out analytically, we take the  approximation $\left(\frac{\partial
f}{\partial x_i}\right)^2 (x_i-\bar{x}_i)^2\approx
\left[f(\bar{x}_i\pm \Delta x_i)-f(\bar{x}_i)\right]^2$ in the
numerical calculations.

\begin{figure}
 \centering
 \includegraphics[totalheight=4cm,width=5cm]{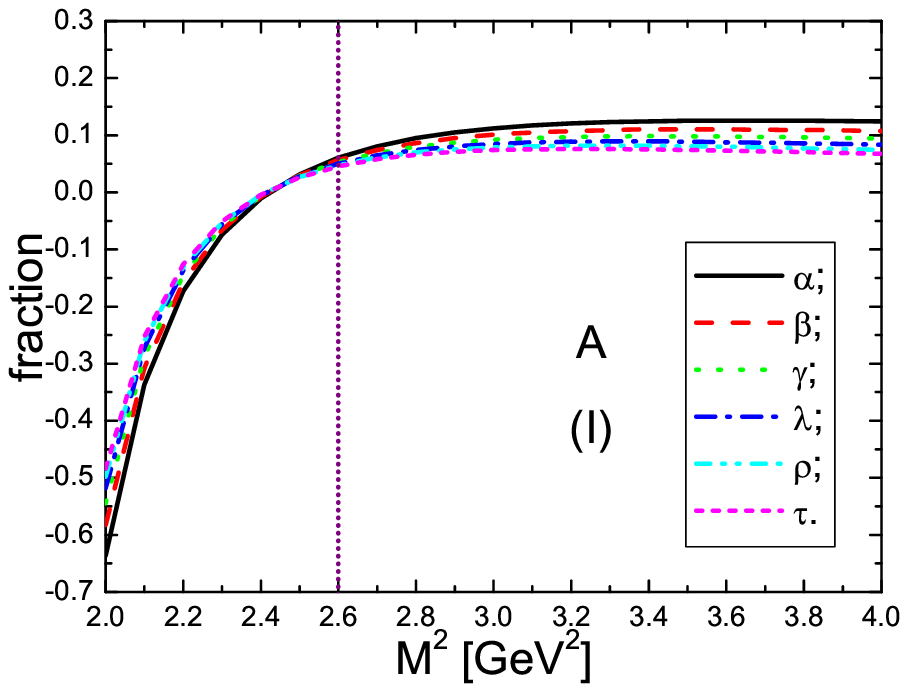}
 \includegraphics[totalheight=4cm,width=5cm]{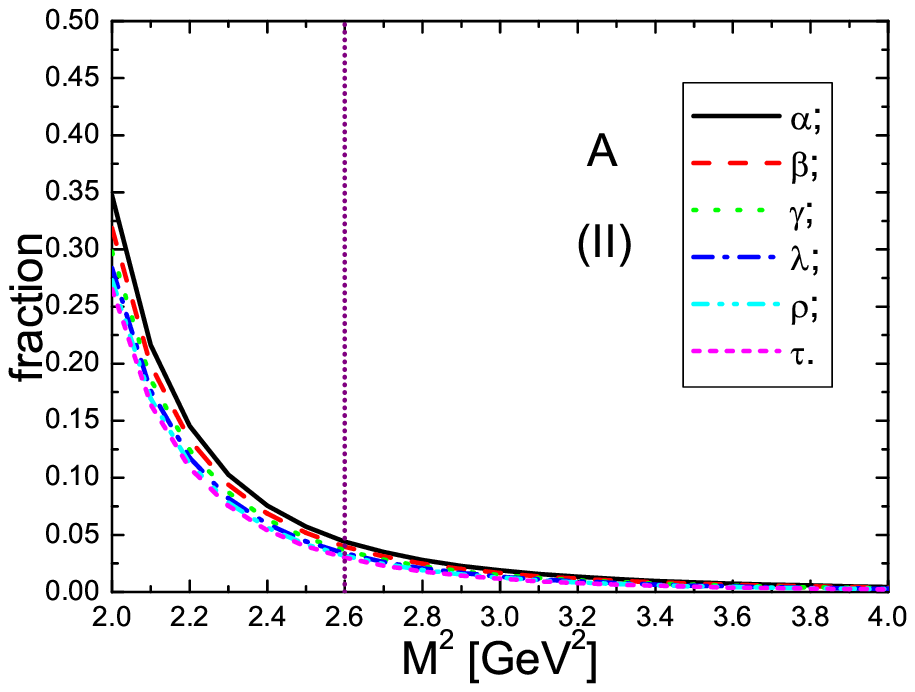}
 \includegraphics[totalheight=4cm,width=5cm]{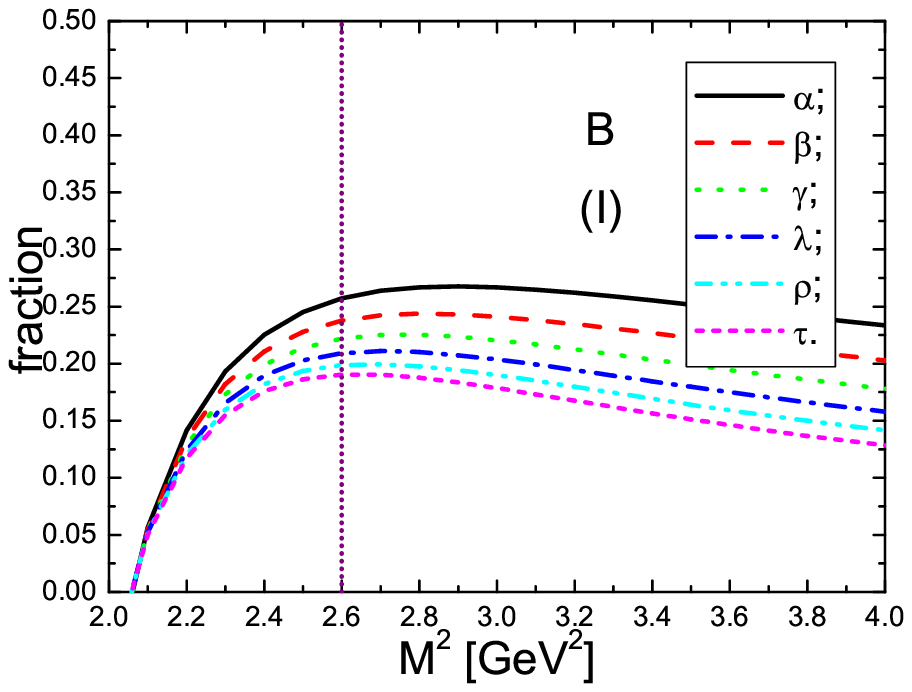}
 \includegraphics[totalheight=4cm,width=5cm]{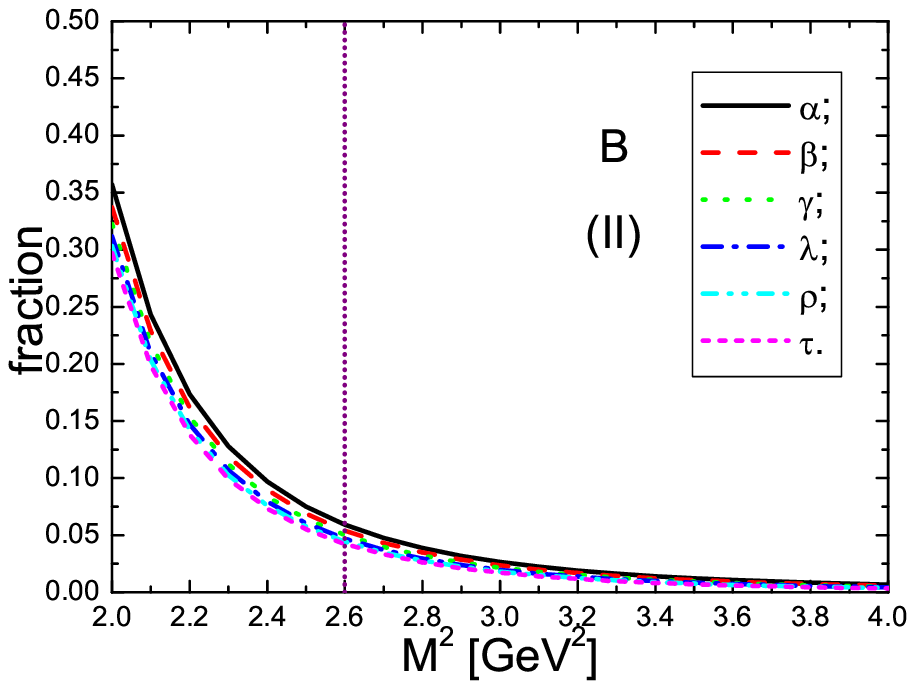}
 \includegraphics[totalheight=4cm,width=5cm]{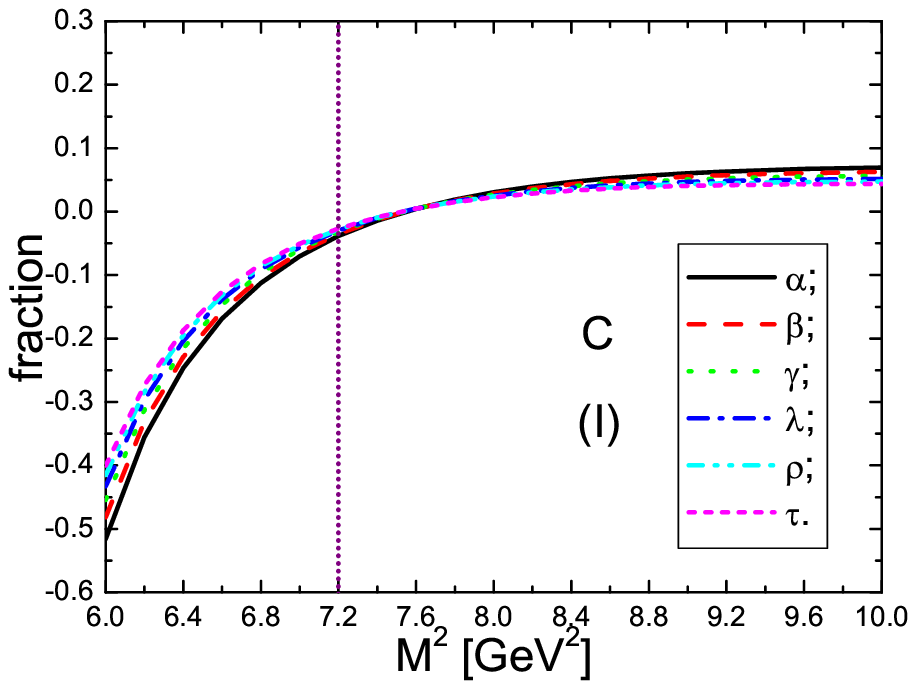}
 \includegraphics[totalheight=4cm,width=5cm]{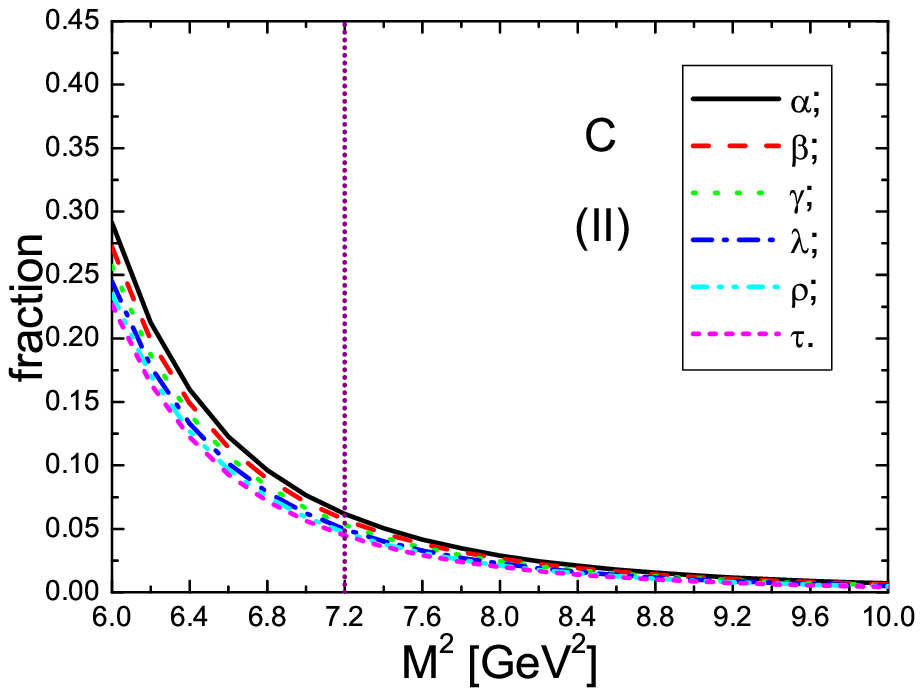}
 \includegraphics[totalheight=4cm,width=5cm]{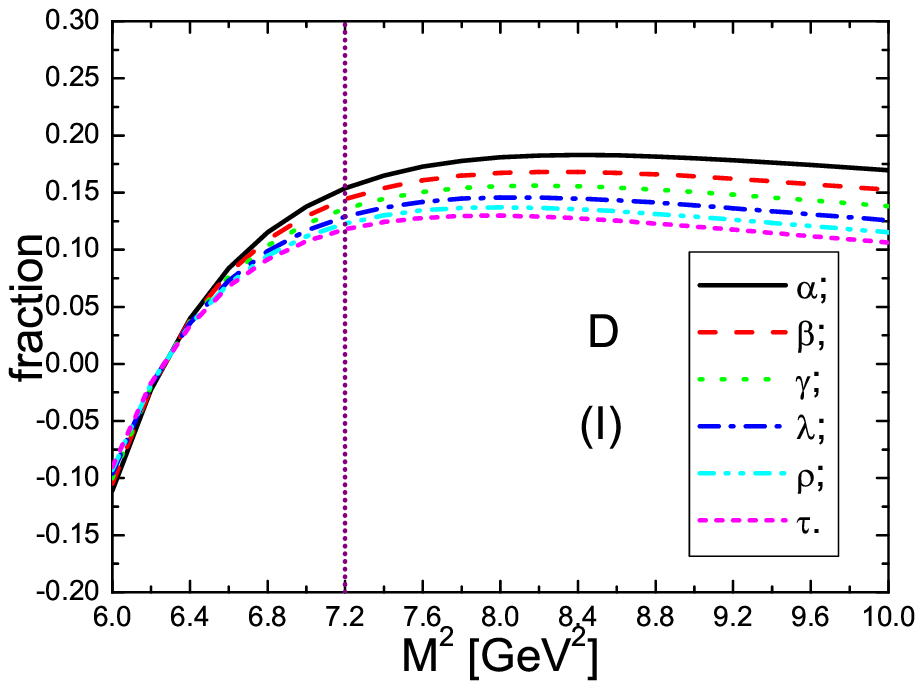}
 \includegraphics[totalheight=4cm,width=5cm]{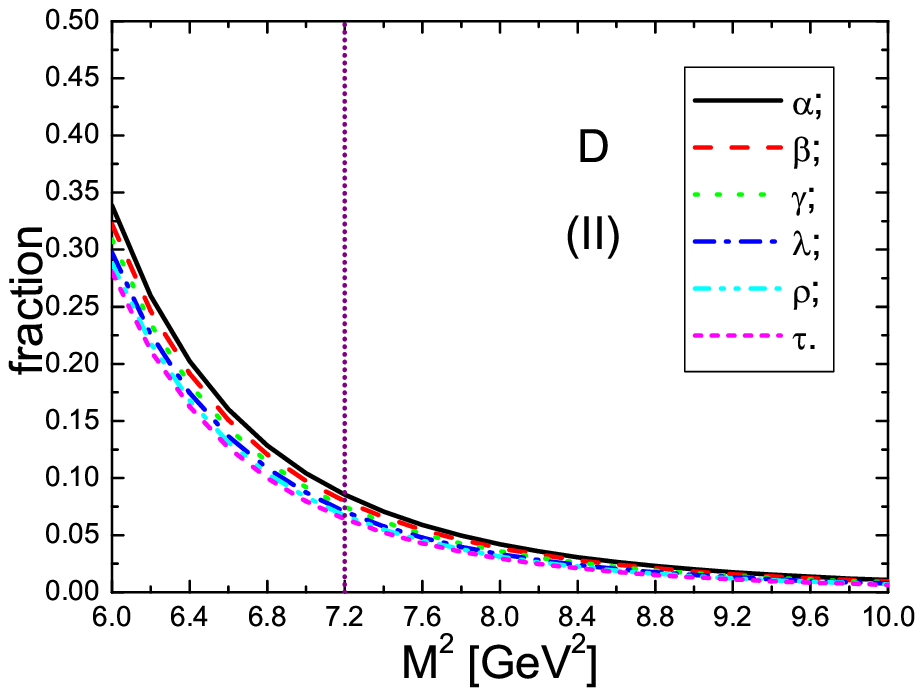}
    \caption{ The contributions from different terms with variation of the Borel
   parameter $M^2$  in the operator product expansion. The (I)  and (II) denote the contributions from the
   $\langle \bar{s}s\rangle^2+\langle \bar{s}s\rangle \langle \bar{s}g_s \sigma Gs\rangle
   $ term and the  $ \langle \bar{s}g_s \sigma Gs\rangle^2
   $ term  respectively.  The $A$, $B$, $C$   and $D$ denote the channels $cc\bar{s}\bar{s}$,
    $cc\bar{q}\bar{q}$,  $bb\bar{s}\bar{s}$ and $bb\bar{q}\bar{q}$ respectively. The notations
   $\alpha$, $\beta$, $\gamma$, $\lambda$,   $\rho$ and $\tau$ correspond to the threshold
   parameters $s_0=22\,\rm{GeV}^2$,
   $23\,\rm{GeV}^2$, $24\,\rm{GeV}^2$, $25\,\rm{GeV}^2$, $26\,\rm{GeV}^2$ and $27\,\rm{GeV}^2$
   respectively in the doubly charmed  channels; while they correspond to the threshold
   parameters $s_0=134\,\rm{GeV}^2$,
   $136\,\rm{GeV}^2$, $138\,\rm{GeV}^2$, $140\,\rm{GeV}^2$, $142\,\rm{GeV}^2$ and $144\,\rm{GeV}^2$ respectively
   in the doubly bottom channels.}
\end{figure}

\begin{figure}
 \centering
   \includegraphics[totalheight=5cm,width=6cm]{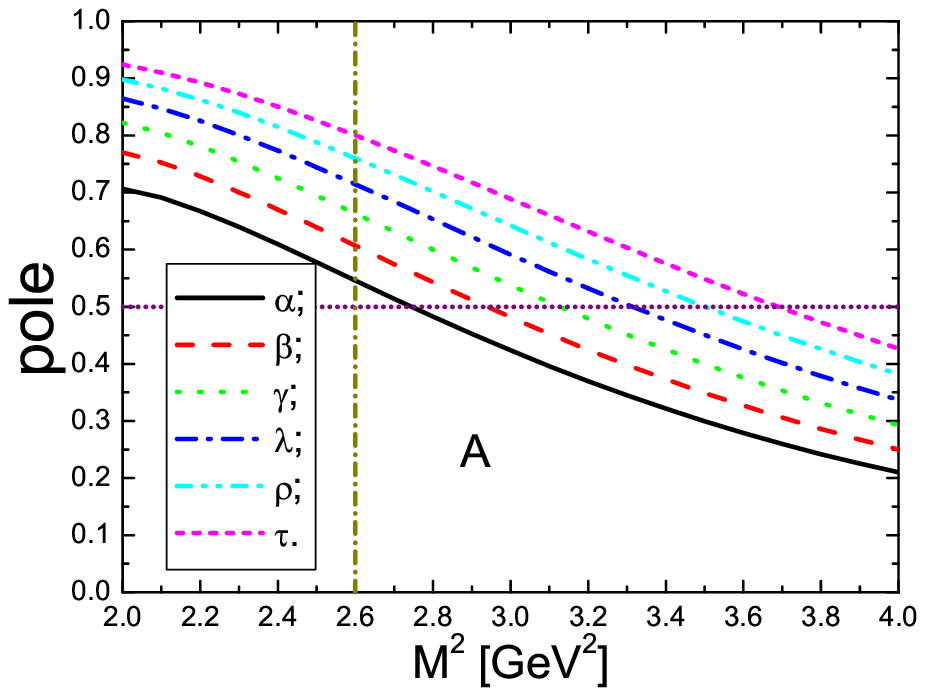}
    \includegraphics[totalheight=5cm,width=6cm]{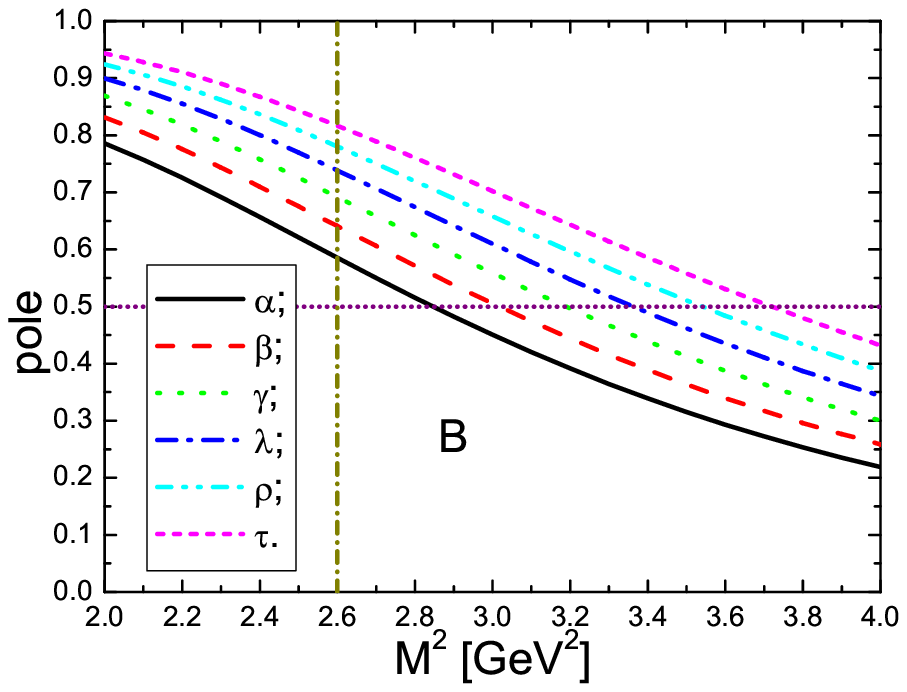}
    \includegraphics[totalheight=5cm,width=6cm]{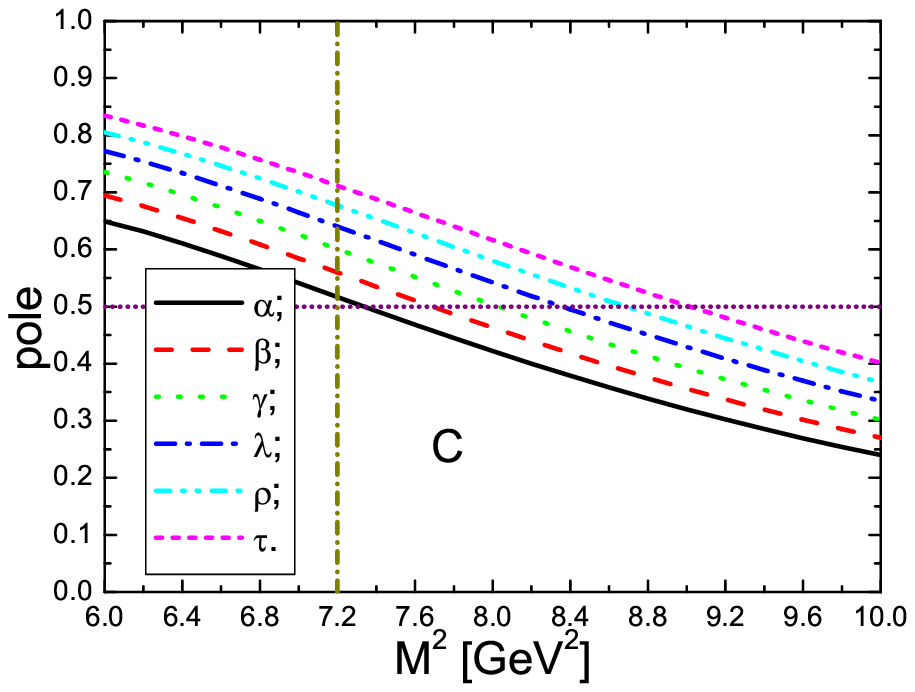}
    \includegraphics[totalheight=5cm,width=6cm]{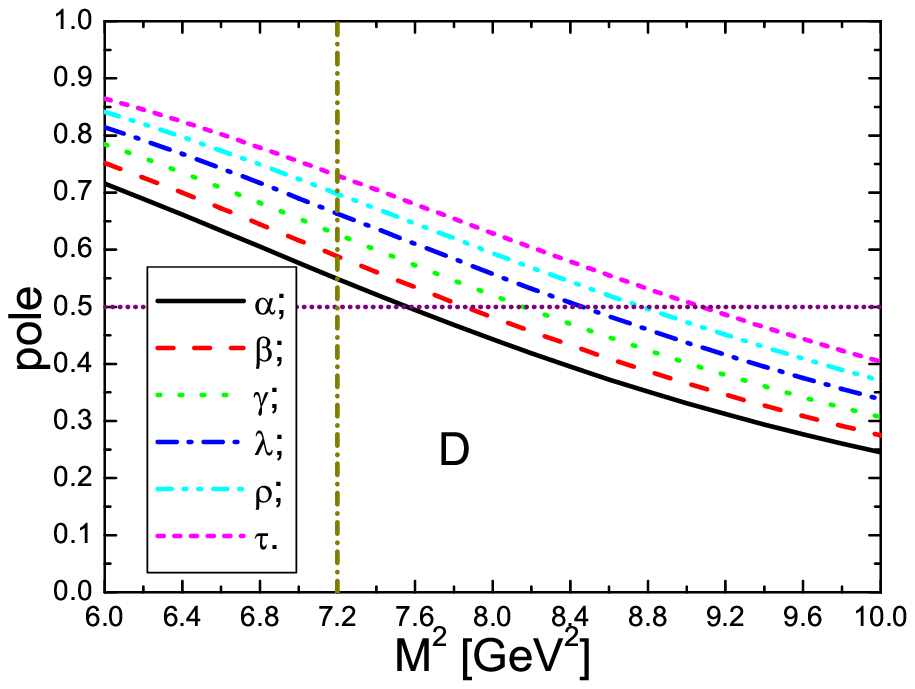}
   \caption{ The contributions of the pole terms with variation of the Borel parameter
   $M^2$. The $A$, $B$, $C$   and $D$ denote the channels $cc\bar{s}\bar{s}$,
    $cc\bar{q}\bar{q}$,  $bb\bar{s}\bar{s}$ and $bb\bar{q}\bar{q}$ respectively. The notations
   $\alpha$, $\beta$, $\gamma$, $\lambda$,   $\rho$ and $\tau$ correspond to the threshold
   parameters $s_0=22\,\rm{GeV}^2$,
   $23\,\rm{GeV}^2$, $24\,\rm{GeV}^2$, $25\,\rm{GeV}^2$, $26\,\rm{GeV}^2$ and $27\,\rm{GeV}^2$
   respectively in the doubly charmed channels; while they correspond to the threshold
   parameters $s_0=134\,\rm{GeV}^2$,
   $136\,\rm{GeV}^2$, $138\,\rm{GeV}^2$, $140\,\rm{GeV}^2$, $142\,\rm{GeV}^2$ and $144\,\rm{GeV}^2$ respectively
   in the doubly bottom channels.}
\end{figure}

\begin{figure}
 \centering
   \includegraphics[totalheight=5cm,width=6cm]{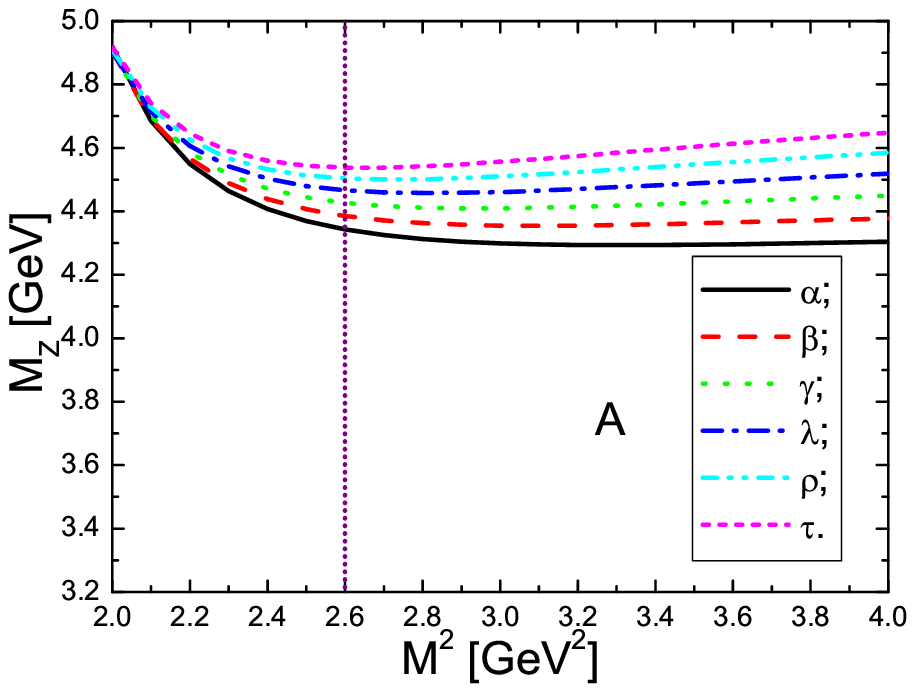}
    \includegraphics[totalheight=5cm,width=6cm]{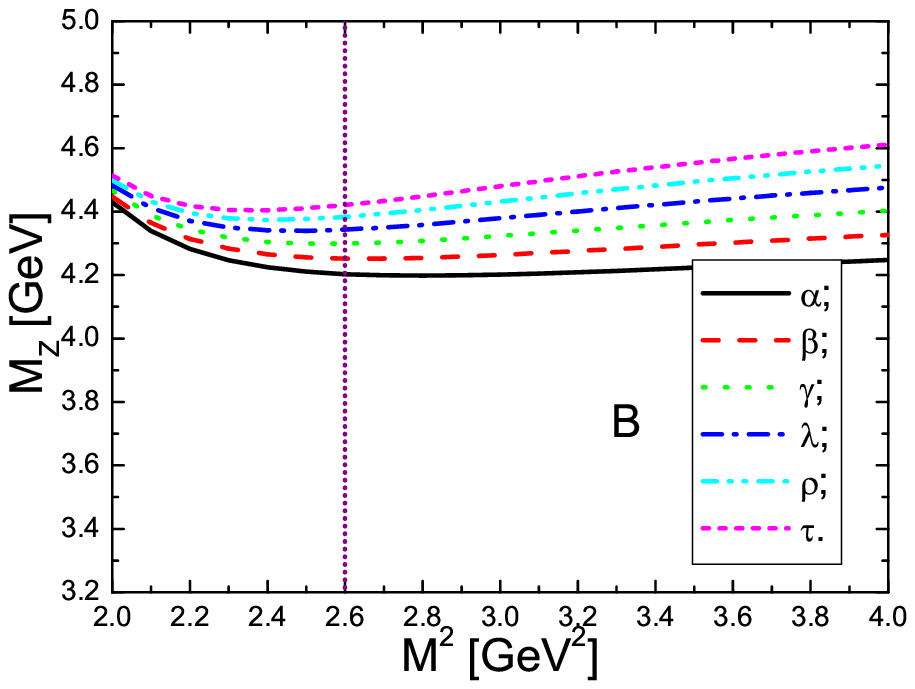}
    \includegraphics[totalheight=5cm,width=6cm]{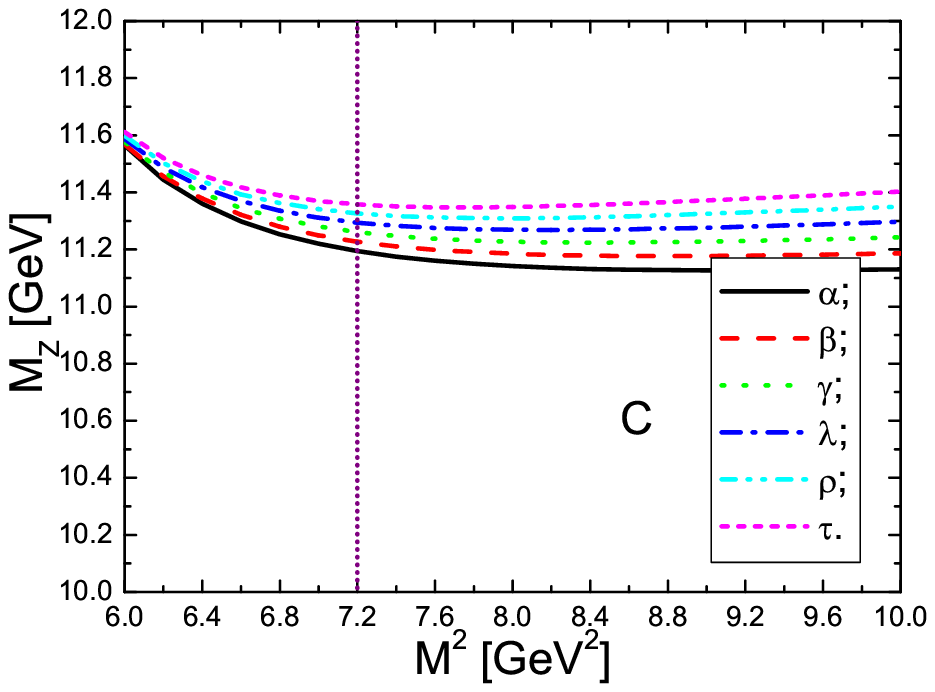}
    \includegraphics[totalheight=5cm,width=6cm]{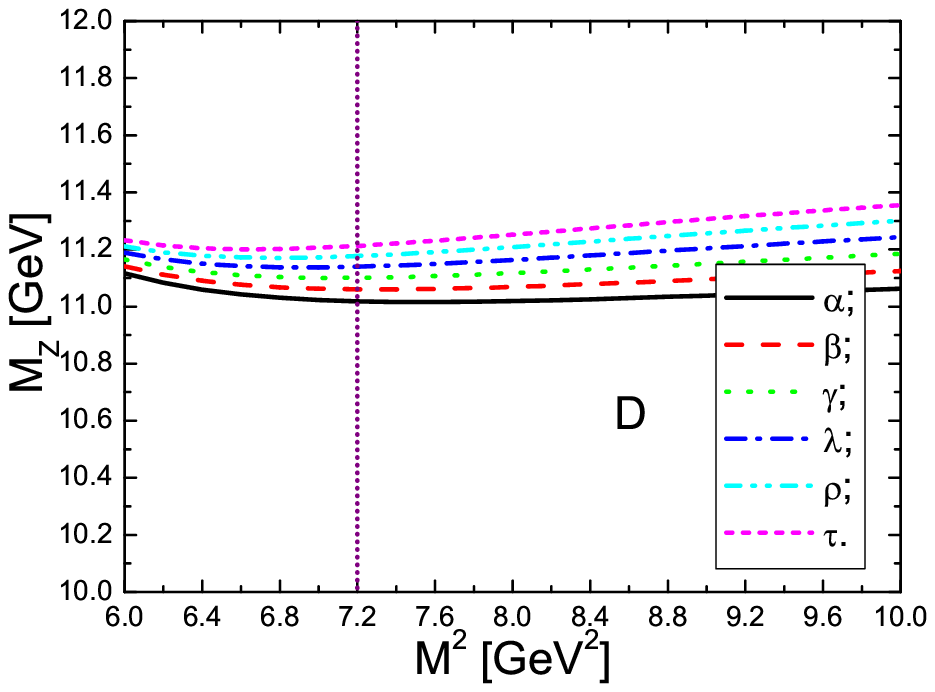}
      \caption{  The masses of the scalar doubly heavy  tetraquark states with variation of the Borel parameter
   $M^2$. The $A$, $B$, $C$   and $D$ denote the channels $cc\bar{s}\bar{s}$,
    $cc\bar{q}\bar{q}$,  $bb\bar{s}\bar{s}$ and $bb\bar{q}\bar{q}$ respectively. The notations
   $\alpha$, $\beta$, $\gamma$, $\lambda$,   $\rho$ and $\tau$ correspond to the threshold
   parameters $s_0=22\,\rm{GeV}^2$,
   $23\,\rm{GeV}^2$, $24\,\rm{GeV}^2$, $25\,\rm{GeV}^2$, $26\,\rm{GeV}^2$ and $27\,\rm{GeV}^2$
   respectively in the doubly charmed channels; while they correspond to the threshold
   parameters $s_0=134\,\rm{GeV}^2$,
   $136\,\rm{GeV}^2$, $138\,\rm{GeV}^2$, $140\,\rm{GeV}^2$, $142\,\rm{GeV}^2$ and $144\,\rm{GeV}^2$ respectively
   in the doubly bottom channels.}
\end{figure}

\begin{figure}
 \centering
    \includegraphics[totalheight=5cm,width=6cm]{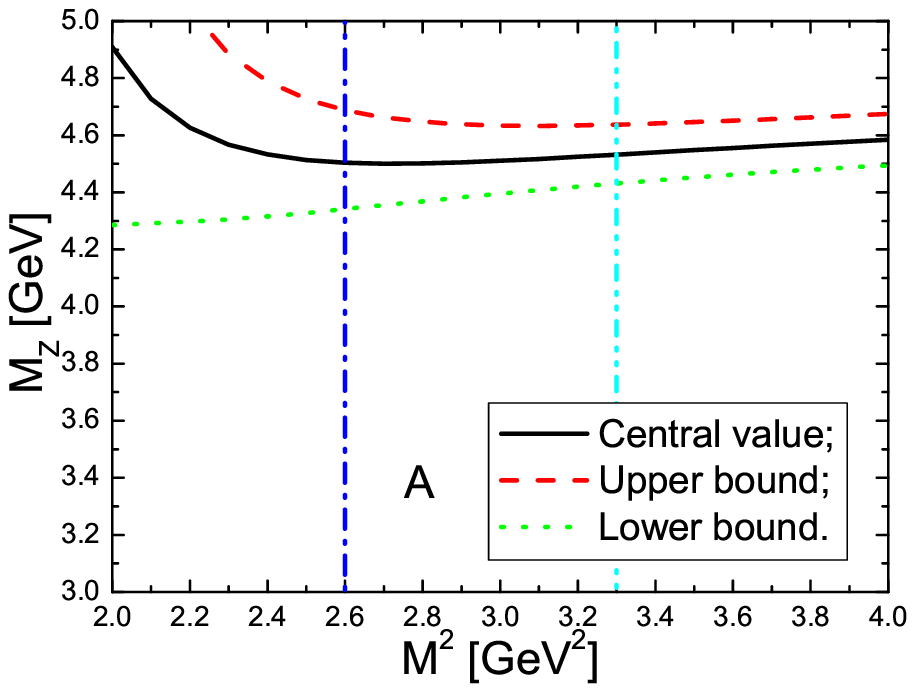}
    \includegraphics[totalheight=5cm,width=6cm]{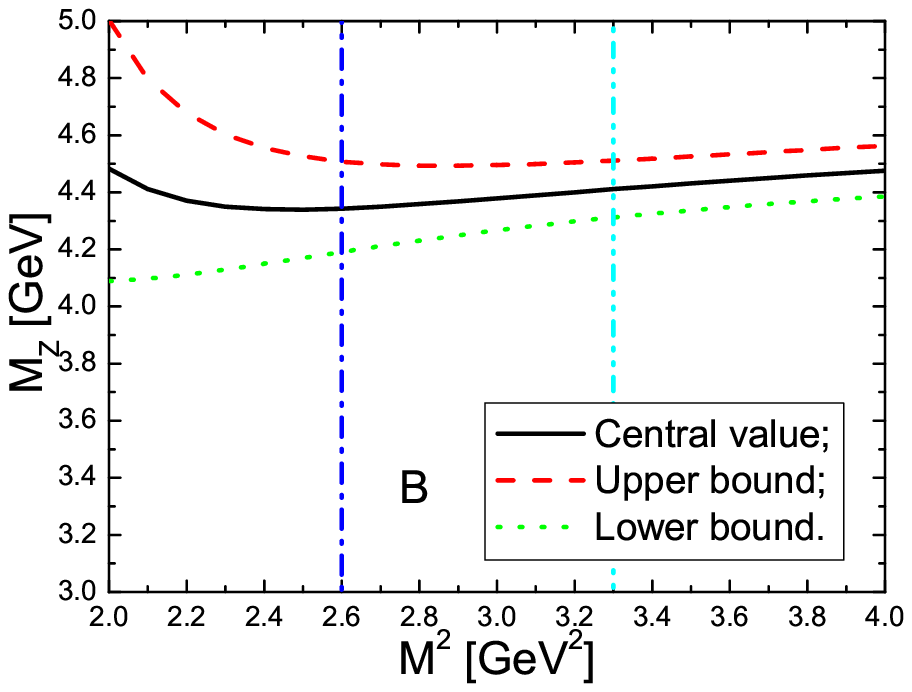}
    \includegraphics[totalheight=5cm,width=6cm]{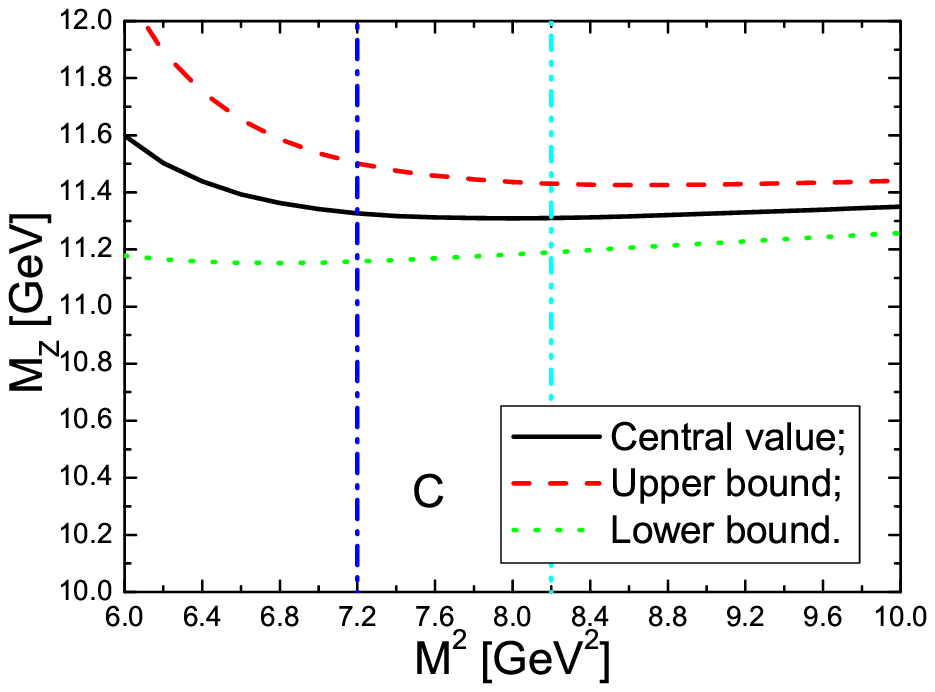}
    \includegraphics[totalheight=5cm,width=6cm]{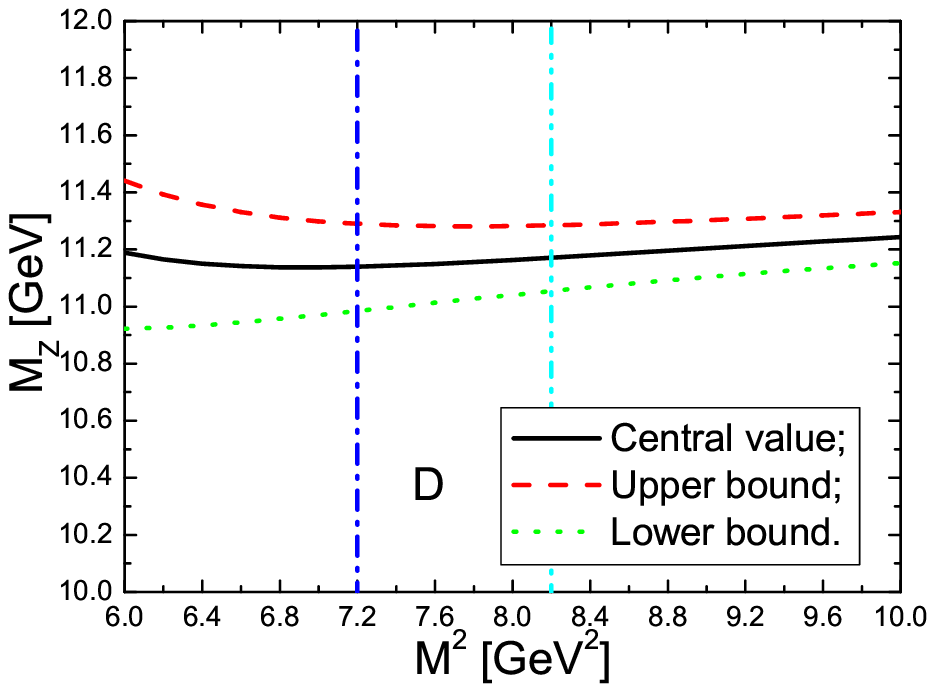}
      \caption{ The masses of the scalar doubly heavy tetraquark states with variation of the Borel parameter
   $M^2$. The $A$, $B$, $C$   and $D$ denote the channels $cc\bar{s}\bar{s}$,
    $cc\bar{q}\bar{q}$,  $bb\bar{s}\bar{s}$ and $bb\bar{q}\bar{q}$ respectively.}
\end{figure}

\begin{figure}
 \centering
    \includegraphics[totalheight=5cm,width=6cm]{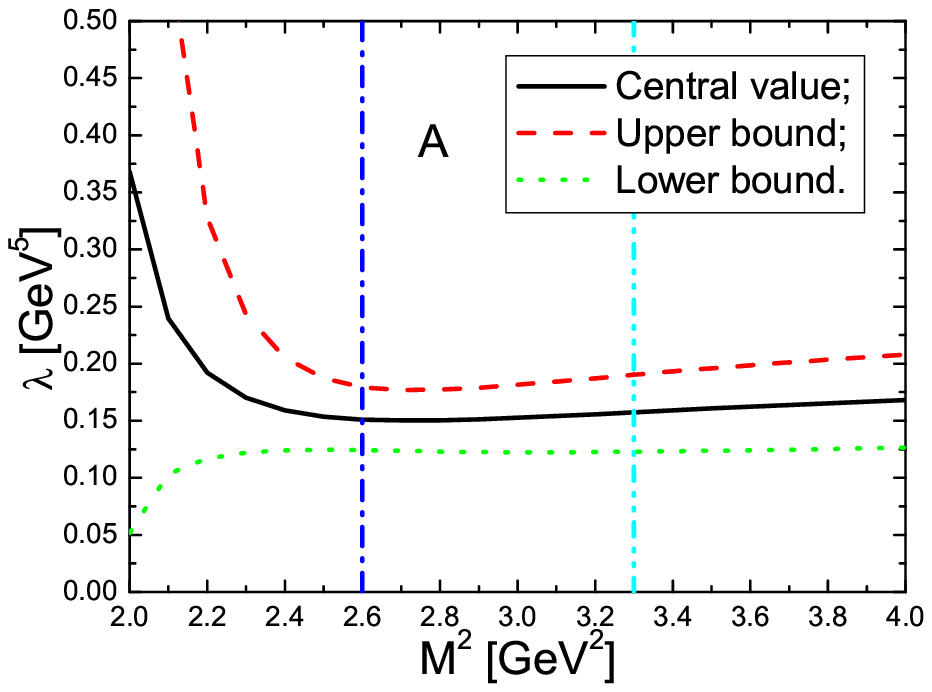}
     \includegraphics[totalheight=5cm,width=6cm]{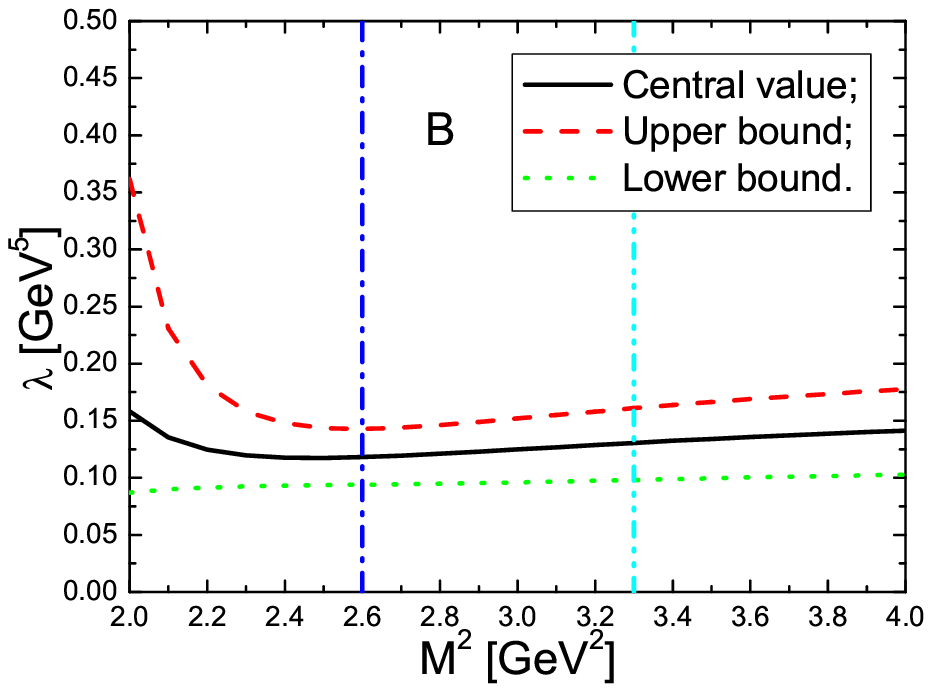}
      \includegraphics[totalheight=5cm,width=6cm]{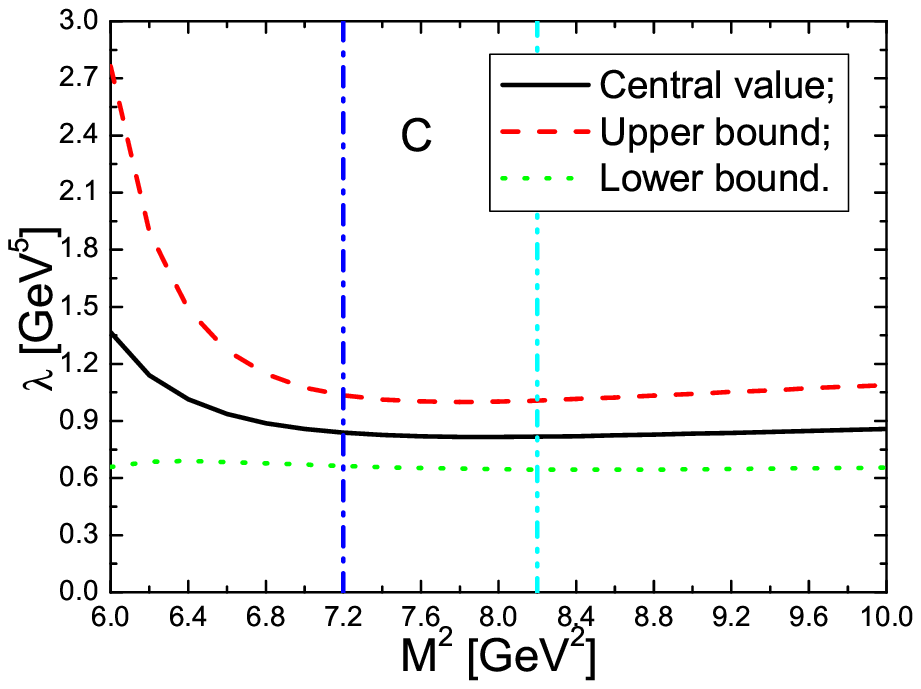}
       \includegraphics[totalheight=5cm,width=6cm]{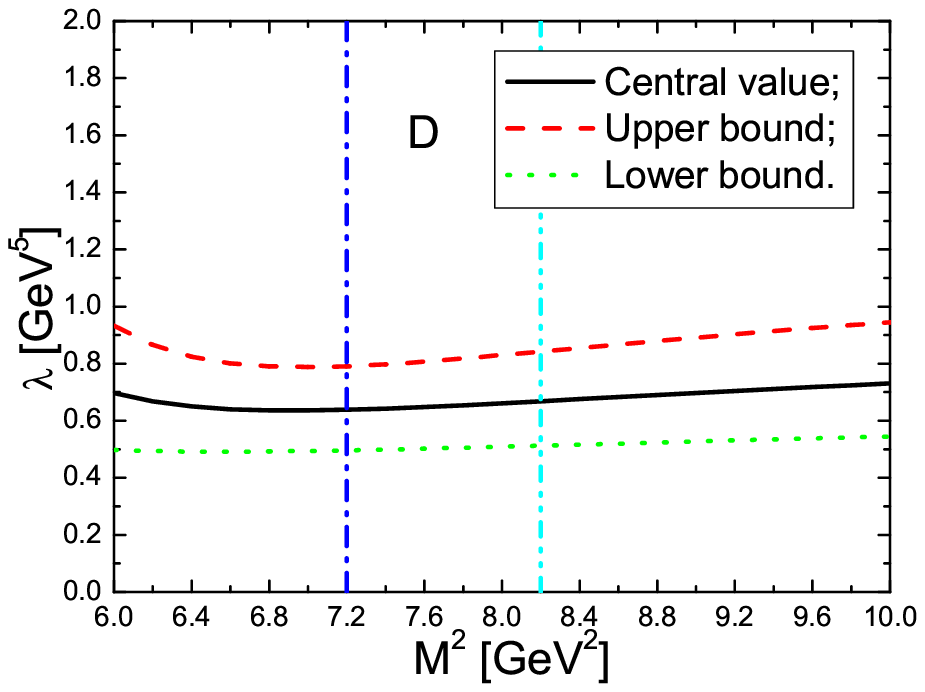}
   \caption{ The pole residues of the scalar doubly heavy tetraquark states with variation of the Borel parameter
   $M^2$. The $A$, $B$, $C$   and $D$ denote the channels $cc\bar{s}\bar{s}$,
    $cc\bar{q}\bar{q}$,  $bb\bar{s}\bar{s}$ and $bb\bar{q}\bar{q}$ respectively.}
\end{figure}

\begin{table}
\begin{center}
\begin{tabular}{|c|c|c|c|}
\hline\hline tetraquark states & $M_Z$ &  $\lambda_Z$ \\
\hline
  $cc\bar{s}\bar{s} $ &$4.52\pm0.18$&  $0.154\pm0.032$ \\      \hline
 $cc\bar{q}\bar{q}$ &$4.35\pm0.16$&  $0.126\pm0.033$ \\ \hline
  $bb\bar{s}\bar{s} $ &$11.32\pm0.18$&  $0.825\pm0.180$ \\ \hline
  $bb\bar{q}\bar{q} $ &$11.14\pm0.16$&  $0.660\pm0.164$ \\ \hline
    \hline
\end{tabular}
\end{center}
\caption{ The masses and the pole residues  of the scalar doubly
heavy tetraquark states. The masses are in unit of $\rm{GeV}$ and
the pole residues are in unit of $\rm{GeV}^5$. }
\end{table}

\begin{table}
\begin{center}
\begin{tabular}{|c|c|c|c|c|c|c|}
\hline\hline  & This work &  \cite{Ebert-2007}& \cite{WangScalar-2}$^*$& \cite{Ebert-ccbar,Ebert-bbbar}$^*$& \cite{CDM-1,CDM-2,CDM-3}$^*$ \\
\hline
  $cc\bar{s}\bar{s} $ &$4.52\pm0.18$&  $4.359$ & $4.45\pm0.16$&$3.967$ & $3.927$\\      \hline
 $cc\bar{q}\bar{q}$ &$4.35\pm0.16$&  $4.056$ &$4.36\pm0.18$ & $3.852$ & $3.832$\\ \hline
  $bb\bar{s}\bar{s} $ &$11.32\pm0.18$&  $10.932$& $11.23\pm0.16$& $10.671$ & $10.874$ \\ \hline
  $bb\bar{q}\bar{q} $ &$11.14\pm0.16$&  $10.648$ &$11.14\pm0.19$& $10.473$ & $10.528$\\ \hline
    \hline
\end{tabular}
\end{center}
\caption{ The masses  of the scalar doubly heavy tetraquark states,
the star denotes the corresponding $Qq\bar{Q}\bar{q}'$ type
tetraquark states. The masses are in unit of $\rm{GeV}$. }
\end{table}

Naively, we expect the $QQ\bar{q}\bar{q}'$ and $Qq\bar{Q}\bar{q}'$
type tetraquark states have degenerate masses as the color
interactions are flavor blinded.  However, we cannot obtain a
relation between the corresponding interpolating currents by Fierz
reordering in the color and Dirac spinor  spaces as different flavor
structures are concerned.  Furthermore, additional contributions
from the instanton configurations make the situation more
complicated \cite{RILM1994}.
 In Table 2, we also
present the values of the $QQ\bar{q}\bar{q}'$  and
$Qq\bar{Q}\bar{q}'$ type tetraquark states from the relativistic
quark model based on a quasipotential approach in QCD
\cite{Ebert-2007,Ebert-ccbar,Ebert-bbbar}, the constituent diquark
model plus the spin-spin
 interactions \cite{CDM-1,CDM-2,CDM-3}, and the QCD sum rules \cite{WangScalar-2}.
 The QCD sum rules indicate the
   $QQ\bar{q}\bar{q}'$  and
$Qq\bar{Q}\bar{q}'$ type tetraquark states have almost degenerate
masses, while the central values of present predictions  are larger
than the corresponding ones from other theoretical models about
$(0.2-0.7)\, \rm{GeV}$.

In Refs.\cite{Ebert-2007,Ebert-ccbar,Ebert-bbbar}, Ebert et al take
the diquarks as bound states of the two quarks in the color
antitriplet channel, and calculate their mass spectrum using a
Schrodinger type equation, then take the masses of the diquarks  as
the basic input parameters, and study the mass spectrum of the heavy
tetraquark states as bound states of the diquark-antidiquark system.
 In Refs.\cite{CDM-1,CDM-2,CDM-3}, Maiani et al
take the diquarks as the basic constituents, examine the rich
spectrum of the diquark-antidiquark states  with  the constituent
diquark masses and the spin-spin
 interactions, and try to  accommodate some of the newly observed charmonium-like and bottonium-like resonances not
 fitting pure $c\bar{c}$ and $b\bar{b}$ assignment. The predictions depend heavily on  the assumption that the light
 scalar mesons $a_0(980)$ and $f_0(980)$ are tetraquark states,
 the  basic  parameters (constituent diquark masses) are
 estimated thereafter. In the conventional quark models, the constituent quark masses  are
taken as the basic input parameters, and fitted to reproduce the
mass spectra  of the well known  mesons and baryons. However, the
present experimental knowledge about the phenomenological hadronic
spectral densities of the tetraquark states is  rather vague, the
constituent diquark masses and thereafter predictions cannot be
confronted with the experimental data.

If   kinematically allowed, the scalar doubly heavy tetraquark
states $Z$ can decay to the  heavy meson pairs with the
Okubo-Zweig-Iizuka super-allowed "fall-apart" mechanism, i.e.
$QQ\bar{q}\bar{q}\to Q\bar{q}\,Q\bar{q}$ and $QQ\bar{s}\bar{s}\to
Q\bar{s}\,Q\bar{s}$. The thresholds for the $DD$, $D_sD_s$,
$D^*D^*$, $D^*_sD^*_s$, $BB$, $B_sB_s$, $B^*B^*$ and $B^*_sB^*_s$
are about $3.74\,\rm{GeV}$, $3.94\,\rm{GeV}$, $4.02\,\rm{GeV}$,
$4.22\,\rm{GeV}$, $10.56\,\rm{GeV}$, $10.73\,\rm{GeV}$,
$10.65\,\rm{GeV}$ and $10.83\,\rm{GeV}$, respectively \cite{PDG}.
From Table 1, we can see that the strong decays
$Z_{cc\bar{q}\bar{q}} \to DD,\,D^*D^*$ and $Z_{cc\bar{s}\bar{s}} \to
D_sD_s,\,D_s^*D_s^*$ are kinematically allowed, the phase spaces are
rather large; while the corresponding decays  for the doubly bottom
tetraquark states are kinematically forbidden.

The doubly heavy tetraquark states can also decay to  the baryon
pairs with creation of the $q\bar{q}$ or $s\bar{s}$ pairs from the
QCD vacuum, $QQ\bar{q}\bar{q} \to
QQq'+\bar{q}'\bar{q}\bar{q},\,QQs+\bar{s}\bar{q}\bar{q}$,
$QQ\bar{s}\bar{s} \to
QQq+\bar{q}\bar{s}\bar{s},\,QQs+\bar{s}\bar{s}\bar{s}$. However, the
strong decays to  $\Xi_{QQ}\bar{p}$,  $\Xi^*_{QQ}\bar{\Delta}$,
$\Omega_{QQ}\bar{\Sigma}$, $\Omega^*_{QQ}\bar{\Sigma}^*$,
$\Xi_{QQ}\bar{\Xi}'$, $\Xi^*_{QQ}\bar{\Xi}^*$,
$\Omega_{QQ}\bar{\Omega}$ and $\Omega^*_{QQ}\bar{\Omega}^*$ are
kinematically forbidden or greatly suppressed.  The scalar doubly
charmed  tetraquark states maybe have large widths, while the scalar
doubly bottom tetraquark states maybe have very small widths.

In 2002, the SELEX collaboration reported the first observation of a
signal for the doubly charm baryon  state $ \Xi_{cc}^+$ in the
charged decay mode $\Xi_{cc}^+\rightarrow\Lambda_c^+K^-\pi^+$
\cite{SELEX2002}, and confirmed later by the same collaboration in
the decay mode $\Xi_{cc}^+\rightarrow pD^+K^- $ with measured mass
$M_{\Xi}=3518.9 \pm 0.9 \,\rm{ MeV }$ \cite{SELEX2004}. No other
doubly heavy baryon states are observed. We use the  masses of the
${1\over 2}^\pm$ and ${3\over2}^\pm$ doubly heavy baryon states
$\Xi_{QQ}$, $\Xi^*_{QQ}$, $\Omega_{QQ}$, $\Omega^*_{QQ}$ from the
QCD sum rules \cite{Wang-QQ1,Wang-QQ2,Wang-QQ3}.

The LHCb is a dedicated $b$ and $c$-physics precision experiment at
the LHC (large hadron collider). The LHC will be the world's most
copious  source of the $b$ hadrons, and  a complete spectrum of the
$b$ hadrons will be available through gluon fusion. In proton-proton
collisions at $\sqrt{s}=14\,\rm{TeV}$¡Ì, the $b\bar{b}$ cross
section is expected to be $\sim 500\mu b$ producing $10^{12}$
$b\bar{b}$ pairs in a standard  year of running at the LHCb
operational luminosity of $2\times10^{32} \rm{cm}^{-2}
\rm{sec}^{-1}$ \cite{LHC}. The scalar doubly heavy tetraquark states
predicted in the present work may be observed at the LHCb, if they
exist indeed.

\section{Conclusion}
In this article, we study the mass spectrum of the scalar doubly
charmed  and doubly bottom tetraquark states  with the QCD sum rules
in a systematic way. The mass spectrum are calculated  by imposing
the two criteria (pole dominance and convergence of the operator
product expansion) of the QCD sum rules. The present predictions can
be confronted with the experimental data in the future at the LHCb.

\section*{Appendix}
The spectral densities $\rho_{Z}(s)$ at the level of the quark-gluon
degrees of freedom:

\begin{eqnarray}
\rho_Z(s)&=&\frac{1}{32 \pi^6} \int_{\alpha_{i}}^{\alpha_{f}}d\alpha
\int_{\beta_{i}}^{1-\alpha} d\beta
\alpha\beta(1-\alpha-\beta)^3(s-\widetilde{m}^2_Q)^2(7s^2-6s\widetilde{m}^2_Q+\widetilde{m}^4_Q)
\nonumber \\
&&+\frac{m_Q^2}{32 \pi^6} \int_{\alpha_{i}}^{\alpha_{f}}d\alpha
 \int_{\beta_{i}}^{1-\alpha} d\beta (1-\alpha-\beta)^2(s-\widetilde{m}^2_Q)^3
\nonumber \\
&&+\frac{ m_s\langle \bar{s}s\rangle}{ \pi^4}
\int_{\alpha_{i}}^{\alpha_{f}}d\alpha \int_{\beta_{i}}^{1-\alpha}
d\beta \alpha \beta (1-\alpha-\beta)(10s^2-12s\widetilde{m}^2_Q+3\widetilde{m}^4_Q)   \nonumber\\
&&-\frac{ m_s\langle \bar{s}s\rangle}{ \pi^4}
\int_{\alpha_{i}}^{\alpha_{f}}d\alpha \int_{\beta_{i}}^{1-\alpha}
d\beta \alpha \beta(s-\widetilde{m}^2_Q)(2s-\widetilde{m}^2_Q)   \nonumber\\
&&-\frac{ m_s\langle \bar{s}g_s\sigma Gs\rangle}{ \pi^4}
\int_{\alpha_{i}}^{\alpha_{f}}d\alpha \int_{\beta_{i}}^{1-\alpha}
d\beta \alpha \beta
\left[2s-\widetilde{m}^2_Q+\frac{s^2}{6}\delta(s-\widetilde{m}^2_Q)\right]   \nonumber\\
&& -\frac{3 m_sm_Q^2\langle \bar{s}s\rangle}{2 \pi^4}
\int_{\alpha_{i}}^{\alpha_{f}}d\alpha \int_{\beta_{i}}^{1-\alpha}
d\beta(s-\widetilde{m}^2_Q)  \nonumber\\
&&+\frac{4 m_Q^2  \langle \bar{s}s\rangle^2}{3 \pi^2}
\int_{\alpha_{i}}^{\alpha_{f}} d\alpha +\frac{2   \langle
\bar{s}s\rangle^2}{3 \pi^2} \int_{\alpha_{i}}^{\alpha_{f}} d\alpha
\alpha(1-\alpha)(3s-2\widetilde{m}^2_Q)\nonumber\\
&&+\frac{ m_s\langle \bar{s}g_s\sigma Gs\rangle}{4 \pi^4}
\int_{\alpha_{i}}^{\alpha_{f}}d\alpha \alpha (1-\alpha)
(3s-2\widetilde{m}^2_Q)   \nonumber\\
&&+\frac{5m_sm_Q^2\langle \bar{s}g_s\sigma Gs\rangle }{12 \pi^4}
\int_{\alpha_{i}}^{\alpha_{f}} d\alpha  \nonumber\\
&&-\frac{2m_Q^2 \langle \bar{s}s\rangle\langle \bar{s}g_s \sigma
Gs\rangle }{3 \pi^2} \int_{\alpha_{i}}^{\alpha_{f}} d\alpha
\left[1+\frac{s}{M^2} \right]\delta(s-\widetilde{\widetilde{m}}^2_Q)\nonumber\\
&&-\frac{ \langle \bar{s}s\rangle\langle \bar{s}g_s \sigma Gs\rangle
}{ \pi^2} \int_{\alpha_{i}}^{\alpha_{f}} d\alpha \alpha(1-\alpha)
\left\{2+\left[\frac{4s}{3}+\frac{s^2}{3M^2} \right]\delta(s-\widetilde{\widetilde{m}}^2_Q)\right\}\nonumber\\
&&+\frac{ m_Q^2  \langle \bar{s}g_s \sigma Gs\rangle^2 }{12 \pi^2
M^6} \int_{\alpha_{i}}^{\alpha_{f}} d\alpha
s^2\delta(s-\widetilde{\widetilde{m}}^2_Q)\nonumber\\
&&+\frac{\langle \bar{s}g_s \sigma Gs\rangle^2 }{4 \pi^2}
\int_{\alpha_{i}}^{\alpha_{f}} d\alpha \alpha(1-\alpha)
\left[1+\frac{s}{M^2}
+\frac{s^2}{2M^4}+\frac{s^3}{6M^6}\right]\delta(s-\widetilde{\widetilde{m}}^2_Q)\,
,
\end{eqnarray}
where $\alpha_{f}=\frac{1+\sqrt{1-4m_Q^2/s}}{2}$,
$\alpha_{i}=\frac{1-\sqrt{1-4m_Q^2/s}}{2}$, $\beta_{i}=\frac{\alpha
m_Q^2}{\alpha s -m_Q^2}$,
$\widetilde{m}_Q^2=\frac{(\alpha+\beta)m_Q^2}{\alpha\beta}$,
$\widetilde{\widetilde{m}}_Q^2=\frac{m_Q^2}{\alpha(1-\alpha)}$, and
$\Delta_Z=4(m_Q+m_s)^2$.

\section*{Acknowledgements}
This  work is supported by National Natural Science Foundation of
China, Grant Numbers 10775051, 11075053, and Program for New Century
Excellent Talents in University, Grant Number NCET-07-0282, and the
Fundamental Research Funds for the Central Universities.

\end{document}